# Tailoring Near-Field−Mediated Photon Electron Interactions with Light Polarization

Fatemeh Chahshouri[1,*] and Nahid Talebi[1,2,*]

[1]*Institute of Experimental and Applied Physics, Kiel University, 24098 Kiel, Germany*
[2]*Kiel, Nano, Surface, and Interface Science − KiNSIS, Kiel University, 24098 Kiel, Germany*

E-Mail: talebi@physik.uni-kiel.de; Chahshuri@physik.uni-kiel.de

**Abstract –**

Inelastic interaction of free-electrons with optical near fields has recently attracted attention for manipulating and shaping free-electron wavepackets. Understanding the nature and the dependence of the inelastic cross section on the polarization of the optical near-field is important for both fundamental aspects and the development of new applications in quantum-sensitive measurements. Here, we investigate the effect of the polarization and the spatial profile of plasmonic near-field distributions on shaping free-electrons and controlling the energy transfer mechanisms, but also tailoring the electron recoil. We particularly show that polarization of the exciting light can be used as a control knop for disseminating the acceleration and deceleration path ways via the experienced electron recoil. We also demonstrate the possibility of tailoring the shape of the localized plasmons by incorporating specific arrangements of nanorods to enhance or hamper the transversal and longitudinal recoils of free-electrons. Our findings open up a route towards plasmonic near-fields-engineering for the coherent manipulation and control of slow electron beams for creating desired shapes of electron wavepackets.

## Introduction

By the early 21$^{th}$ century, research into photon-induced near-field electron microscopy (PINEM)[1] opens exciting new techniques in manipulating and shaping[2,3] electron wavepackets. In PINEM, electron wavepackets interact with laser-induced near-field excitations, allowing for spatially, energetically, and temporally resolving the optical modes of nanostructures and materials excitations, such as phonon polaritons[4]. Coherent control of the shape of quantum wavepackets has applications in bond-selective chemistry[5], quantum computing[6,7], and ultrafast control of plasmonic systems[8]. By developing the pump-probe electron spectroscopy and its theoretical quantum description[9–11], investigating the ultrafast dynamics of quantum systems with energetic free electrons and femtosecond light pulses have experienced an impressive boost toward shaping the electron beams[12,13]. Modulating the continuous energy spectrum of free-electron wavepackets [8,14–16] with the evanescent field[17] and bunching the free-electron wavepacket with light[17,18] has promises in quantum technology and sensing. This quantized energy-momentum exchange in electron-photon coupling offers a new degree of freedom for nano-scale spectroscopy, attosecond control of free-electron quantum wavepackets and for temporal electron pulse manipulation[19,20]. In addition to the longitudinal spread of the electron momentum distribution [21], the wiggling motion of electron wavepackets in the electromagnetic field leads to a transversal Lorentz force as well and, consequently, time-dependent vertical elastic deflection of the electron[20,22–25]. In addition, quantum mechanical diffraction that leads to distinct diffraction orders are observed. This effect paves the way toward ultrafast electron diffraction experiments and manipulating the spatial profile of the electron beam using near-field light as well.

The near-field zone that is responsible for inelastic interactions is tightly bound to the nanoparticle surface for slow electrons[26]. Consequently, electron wavepackets are effectively shaped according to the geometry of the nanoparticle[26]. For slow electrons, the inelastic interaction of optical near fields with electrons take advantages from the long interaction time and thus the experienced electron recoil is enhanced. Therefore, when the momentum matching is achieved over a long interaction region and duration, the electron recoil can be boosted and tailored. Furthermore, designing shapes[27], sizes[1], and geometric configurations[28] of metallic (plasmonic) nanostructures, but also laser beam properties such as wavelength and incident polarization angle[21] can provide a machinery for generating intense and highly localized plasmonic confined near-field modes[17] to foster synchronous motions between near-field oscillations and electron wavepacket – thus acting as a designing platform for shaping the free-electron wavepackets in both longitudinal and transverse directions.

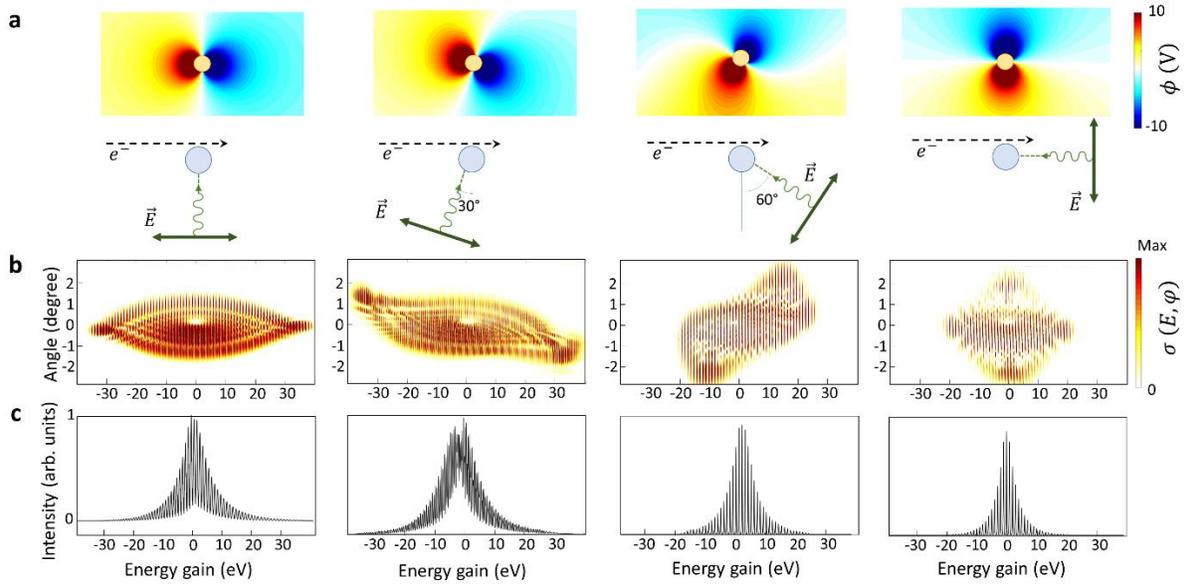

**Figure 1.** Polarization-dependent energy and momentum exchange between laser-induced near-field plasmons of a gold nanorod and an electron wavepacket. (a) Schematics of different inclination angles and polarizations of the incoming laser field. (b) Inelastic scattering cross section, and (c) PINEM spectra for each interaction scenario shown in (a). The kinetic energy of the electron beam is 1 keV, and its FWHM longitudinal and transverse broadenings are 40 nm and 5 nm, respectively. The centre wavelength of the laser pulse is $\lambda = 800$ nm, its temporal broadening is 5 fs, and its peak field amplitude is 5 GVm$^{-1}$.

Here we numerically investigate the interaction of slow electron wavepackets with localized plasmons. Particularly, we intend to explore the experienced transverse recoil of the electron beam and its dependence to the polarization of the incident laser beam interacting with plasmonic nanoparticles. We define the polarization-dependent optical near-field excitations as a quantity for populating selective momentum orders. We also demonstrate that we can pattern the final spatial distribution of the electron beam by tuning the gap configuration of plasmonic nanostructures and thus shaping the plasmonic near-field modes. We further show that the spacing between longitudinal attosecond bunches and transversal momentum exchange orders can be controlled by shaped plasmonic nanogaps.

**Results and Discussions**

The interaction of electron wavepackets with laser-induced plasmon excitations results in an ultrafast amplitude and phase modulation and consequently, the energy modulation and also diffraction of electron wavepackets. The relevant quantity to calculate is thus the angle (momentum)-resolved inelastic cross-section, as we outline later. First, we investigate such effects in two-dimensional space when specific plasmonic modes of a gold nanorod are considered. Moreover, we analyse the effect of the polarization of the incident laser beam on the experienced energy modulation and recoil of the electron wavepacket. Here, we employ our recently developed self-consistent Maxwell-Schrödinger numerical toolbox[9,22] to investigate dynamics of the spatial and spectral evolution of the electron pulse by the laser-induced near-field light. In this semi-classical approach, Maxwell's and Schrödinger equations are combined in a time-dependent loop, using the minimal coupling Hamiltonian[26]. As a result, after the interaction, the final electron wavepacket ($\psi_f(\vec{r}, t)$) is analysed to visualize the dynamics of the electron recoil and energy transfer from the electromagnetic field to the electron wavepacket.

The energy modulation of the electron wavepacket is given by solving the expectation value of the electron kinetic-energy operator as:

$$\langle \psi(x, y; t \to \infty) | \hat{H} | \psi(x, y; t \to \infty) \rangle = \frac{\hbar^2}{2m_0} \iint dk_x dk_y (k_x^2 + k_y^2) |\tilde{\psi}(k_x, k_y; t \to \infty)|^2, \quad (1)$$

where $(x, y)$ and $(k_x, k_y)$ in equation (1) denote the real and reciprocal space coordinates, and $\tilde{\psi}(k_x, k_y; t \to \infty)$ is the Fourier transform of the wave function after the interaction.

In the one-dimensional space, when the electron recoil is neglected, an analytical formalism is obtained to describe the energy transfer along the longitudinal direction. The discrete probability amplitude for the exchange of $n$ quanta of energy between the electron wavepacket and the near-field light are given by expanding the wave function versus a Bessel series using Jacobi-Anger relation[26], as $\psi_n(x, t)^2 \propto J_n^2(2|g|)$, where $J_n$ is the $n$th Bessel function of the first kind, and $g$ is the PINEM coupling strength specified by

$$g = (e/\hbar\omega_{ph}) \int_{-\infty}^{\infty} dx' \, \tilde{E}_x(x', y, \omega) e^{-ix'\omega_{ph}/v_e}, \quad (2)$$

for an electron propagating in the x-direction, and within the nonrecoil approximation [1] $\tilde{E}_x$ is the Fourier transform of the x component of the scattered field. Thus, in general, the electron forms an energy comb with the spacing between the peaks ascertained by the exciting photon energy.

We first simulate the interaction of a slow electron wavepacket with the plasmonic mode around a single nanorod when the nanorod is excited with a linearly polarized laser field, with different excitation angles, using the Maxwell-Schrödinger numerical scheme. As schematically depicted in figure 1, an essential parameter for controlling the coupling and the interaction strength is the direction of the polarization of the excitation field with respect to the electron-beam trajectory. The electron wavepacket has an initial Gaussian distribution at the centre kinetic energy of 1 keV, and is excited by a linearly polarized laser pulse with the center wavelength of $\lambda = 800$ nm and temporal FWHM broadening of 5 fs. The polarization and inclination angles are shown in the figure 1(a). The gold nanorod used as the nanoparticle here has a radius of 15 nm.

By assigning the kinetic energy of the electron as $E = \hbar^2(k_x^2 + k_y^2)/2m_0$, and the scattering angle as $\varphi = \tan^{-1}(k_x/k_y)$, the inelastic scattering cross section $\sigma(E, \varphi) = (m_0/\hbar^2)|\tilde{\psi}(E, \varphi; t \to \infty)|^2$ for different inclination angles is obtained and demonstrated in figure 1(b). In general, a spreading of the electron wavepacket in the momentum space along both longitudinal and transverse directions is observed. Thus, the near-field zone acts as a mediator to overcome the phase mismatch between the light and the electron wavepacket and allows for energy and momentum exchange between them. Due to the small size of the nanoparticle, it can only hosts dipolar plasmonic fields. Thus, retardation and also higher-order multipole excitations are negligible. Here, $\theta$ is defined as the polarization angle with respect to the y-axis. By changing $\theta$, the induced dipolar field (figure 1(a)) align itself along the incident electric field polarization and hence the wiggling motion of the electron beam is oriented also along the same direction. This effect is particularly highlighted by demonstrating the dynamics of the electron wavepacket during the interaction (figure 2). Therefore, the different polarization of the electric field of the incident light can excite plasmonic dipoles with an angle $\theta$ relative to the electron propagation direction and causes a rotation of the populated momentum orders. The PINEM spectra as a function of laser incident angle are depicted in figure 1(c). Altering the incident angle from $\theta = 0$ to $\theta = 90°$, a gradual decrease of the PINEM coupling strength (equation (2)) is observed; thus, the extend of energy gain and loss orders is reduced. In contrast, the experienced recoil along the transverse direction is enhanced, and distinct high-angle scattering lobes are observed. For $\theta = 60°$, three regions in the inelastic scattering cross section are distinguished, namely, (i) the domain within the range $-2° < \varphi < 2°$, where the gain and loss peaks are rather symmetrically distributed, (ii) the domain $\varphi < -2°$, where the energy transfer is centered at $-15$ eV and within the range $-20$ eV to $-10$ eV, and (iii) vice versa for $\varphi > 2°$, where the energy gain peaks within the range 10 eV to 20 eV are observed. This obvious asymmetry could allow for selective acceleration and deceleration of the electron beams and their detection using a mechanical slit, controlled via the near-field polarization.

Thus, concomitant electron acceleration and deceleration caused by the rotational wiggling motion of the electron beam in the dipolar near-field distribution, and quantum mechanical gain and loss peaks could be tailored to effectively shape the electron beam and engineer the recoil. For example, by detecting the energy distribution of the electron beam only for $\varphi > 2°$ or $\varphi < -2°$, via tuning the position of a filter like a mechanical slit, a uniform platform for acceleration and deceleration of a bunched electron wavepacket is obtained.

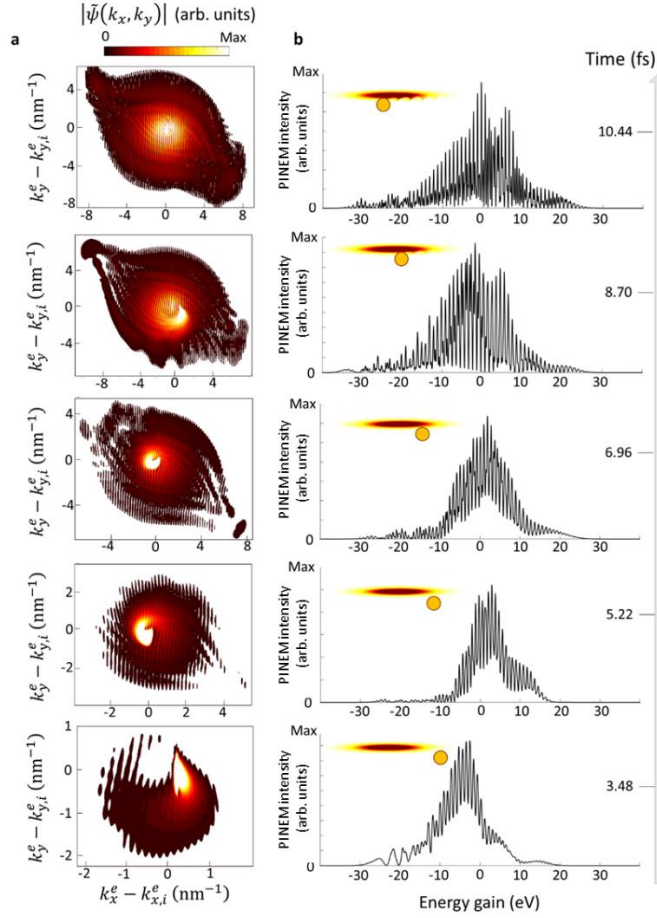

**Figure 2**. Dynamics of the energy and momentum transfer from the near-field to the electron wavepacket during its interaction with localized plasmons. (a) The electron wavepacket in momentum space, and (b) Energy gain spectra calculated during the interaction. The kinetic energy of the electron beam is 1 keV, and its FWHM longitudinal and transverse broadenings are 40 nm and 5 nm, respectively. The centre wavelength of the laser pulse is 800 nm, its temporal broadening is 5 fs, and its peak field amplitude is 5 GVm$^{-1}$. The inclination angle of the laser beam with respect to the electron beam trajectory is 45°.

To better understand the wiggling motion of the electron beam in the optical near-field distribution, we show the dynamics of the electron wavepacket in both momentum and spatial domains, at selected times during the interaction (figure 2). An electron wavepacket at the group velocity of 1 keV interacts with a gold nanorod with the radius of 15 nm, excited by an obliquely polarized light at the center wavelength of $\lambda = 800$ nm and the incidence angle of $\theta = 45°$. Obviously, the extent of the electron wavepacket in the momentum distribution is mostly determined by the spread of the wiggling motion, which is by itself controlled by the classical Lorentz force. The experienced energy gain and loss peaks is semiclassically understood as a phase-modulation phenomena, described with the Volkov representation[9,29]. Thus, the interference between phase-modulated and amplitude modulated waves, appears as distinct energy gain and loss peaks in the momentum distribution of the electron wavepacket.

During the interaction time, the electron experiences either a repelling or attracting force towards the nanostructure, resulting in circular wiggling motions observed in the momentum representation

(figure 2(a)). As shown in figure 2(b), the gradual dynamics of the energy modulations in the scattering of the electron wavepacket off the near-field light can be described as the outcomes of a quantum walk[9,23] in the discrete momentum states spaced by the classical electromagnetic waves. This behavior leads to a gradual occupation of momentum states of the electron wavepacket by the interaction with the near field light. The ability to dynamically control the outcome of the random walk by a few parameters, such as the polarization, and incident angle of the incident optical beams, makes the proposed system as a promising candidate for boson-sampling schemes [30].

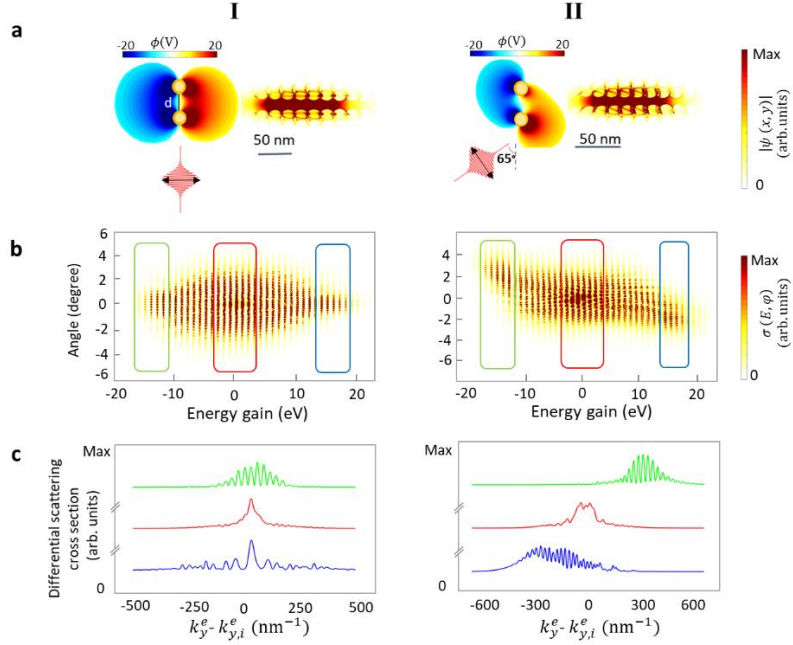

**Figure 3.** Final distribution of the transversal and the longitudinal distributions of free electrons. (I) Normal- and (II) oblique-incident laser pulses. (a) Final modulation of the amplitude of the electron wavepacket in real space, and (b) its inelastic scattering cross section, after its interaction with the enhanced dipolar near-field modes inside the nanogap of a dimer gold nanorod. (c) Electron distribution along the transverse direction after the interaction in the narrow range shown by the highlighted color region in panel b. The electron has the kinetic energy of 200eV ($\beta = 0.028$), and its initial longitudinal and transversal broadenings are 75 nm and 4.5 nm, respectively. Laser field amplitude, wavelength, and temporal broadening are respectively $E_0 = 2$ GV m$^{-1}$, $\lambda = 700$ nm, and 34 fs.

It has been shown elsewhere that an enhanced PINEM coupling strength is achieved by incorporating a so-called phase-matched configuration, where the aforementioned interference paths are constructively accumulated. This could be for example enabled by geometrical considerations, where the oscillation period of the dipole field and the interaction time of the electron beam within the effective interaction length are made to be synchronous[26]. In a system where the phase-matching condition is obtained, the overall recoil that the electron experiences can be additively accumulated as well. We now employ a nanosystem that enables both an enhanced optical near-field by incorporating gap plasmons, as well as a phase-matched configuration, to explore the effect of both on the experienced longitudinal and lateral modulations of the electron wavepacket.

In the following, we investigate the interaction of an electron wavepacket at a kinetic energy of $U_e = 200$ eV with a dimer gold nanorod with a radius of 10 nm and gap spacing of 25 nm. This nanosystem is excited by an x-polarized light (figure 3, I) and a linearly polarized oblique light $\theta = 65°$ (figure 3, II), respectively, both with a center carrier wavelength of $\lambda = 700$ nm and a temporal broadening of 34 fs FWHM, with 2 GVm$^{-1}$ laser field amplitude. The electromagnetic near-field confinement can produce an enhanced field in the middle of the hotspot region in a plasmonic dimer nanostructure with a gap, where the localization of the field and its spatial profile can be controlled by the polarization of the incident field. Thus, owing to the coupling of the dipoles in the nanorods, the coupled plasmonic cavity system can lead to an enhanced near-field localization, which may be more beneficial in shaping electron pulses, as optical field distributions with higher momentum distributions are enabled. This configuration does not allow for a phase-matched interaction. Thus, when only a single nanorod is used, transfer of energy from the near-field to the electron wavepacket, at the extent discussed below are not observed (see supplementary figure 1 for more information).

During the interaction of the electron beam with a plasmonic dimer excited with light, momentum modulation of the electron beam can be controlled by the spatial profile of the electric field (figure 3(a)). In addition to the formation of an attosecond pulse train in real space (figure 3(a)), a quantized modulation of the electron momentum distribution (figure 3(b)) along the longitudinal and transverse directions is observed. Indeed, the gap between the nanorods supports a standing-wave-like pattern within the gap, which can cause the diffraction of the electron wavepacket in a similar way to the kaptitza-Dirac (KD) effect[9,23], while the longitudinal phase modulation results from the already mentioned PINEM effect. As shown in figure 3, for the inelastic scattering cross section for $\theta = 0$, an eye-like pattern is observed in the momentum space (figure 3(b), I). However, for the obliquely excited plasmonic dimer, the populated transverse states are more intensely occupied around the higher order gain and loss peaks (figure 3(b), II). This behavior happens due to the symmetric and asymmetric spatial profile of the localized plasmonic modes with respect to the electron trajectory (figure 3(a)).

The transversal modulation of the electron wavepacket is better visualized by the line profiles shown in figure 3(c). The transverse electron distribution after the interaction, integrated within a short energy range (within the colored boxes depicted in figure 3(b)), shows distinguishable momentum peaks. The incident light with $\theta = 0$ causes a rather symmetric momentum distribution. Because of the overall dipolar pattern of the near-field distribution, the induced near-field at one side of the cavity is against the restoring force acting on the electron at the other side. Therefore, having a symmetric dipolar configuration sets a unified electron transverse recoil in all energetically distributed photon orders. The obliquely excited plasmonic dimer causes an asymmetric recoil distribution over the electron wavepacket. This behavior is much more pronounced for the occupied higher and lower longitudinal momentum states. The summation of the transverse probability distribution within the range of $-4$ eV $\leq E \leq +4$ eV shows a rather symmetric distribution, whereas for the accelerated electrons within the range of $14$ eV $\leq E \leq 20$ eV, or decelerated electron in the range of $-20$ eV $\leq E \leq -14$ eV the interspacing between the maxima of the diffraction orders is roughly equal to $\delta k_y = 30 k_{ph} = 2\pi d^{-1}$ (figure 3(c)).

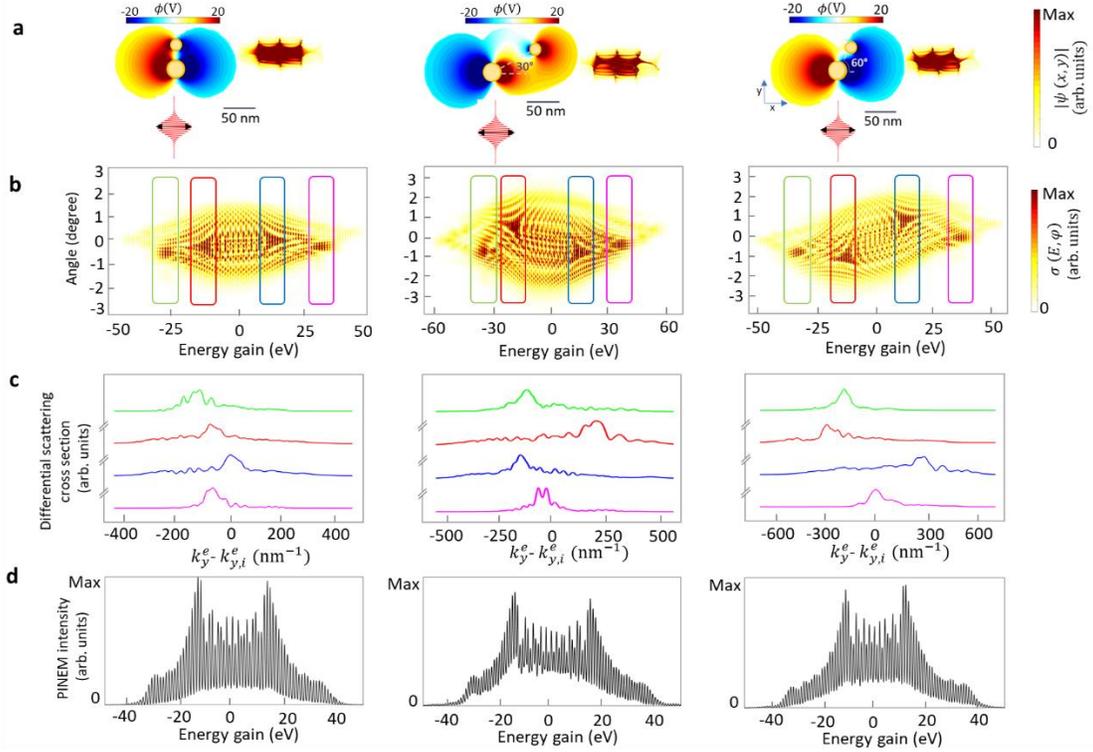

**Figure 4.** Controlling the electron recoil with the near-field distribution of localized plasmons. (I) Both nanorods positioned along the *y*-axis, (II) second nanorod is rotated 30° around the x-axis, and (III) second nanorod is rotated 60° around the x-axis. In all three cases, the project spacing of the nanorods along the y-axis is the same. The amplitude of the electron wavepacket (a) in the real space and (b) its inelastic scattering cross section, after the interaction with the near-field zone excited by the *x*-polarized laser light. The laser pulse has the center wavelength of 700nm, the electric field amplitude of $E_0 = 2$ GVm$^{-1}$, and the temporal FWHM broadening of 34 fs, respectively. (c) Transversal momentum distribution of the electron wavepacket after the interaction with the plasmonic near-field in the specific $k_x$ ranges highlighted in panel b, and (d) PINEM spectra in the overall $k_y$ range. The electron kinetic energy is 600 eV ($\beta = 0.0484c$), and its initial longitudinal and transversal broadenings are 45 nm and 3 nm, respectively.

A more complicated electron-bunching effect is observed by keeping the parameters of the *x*-polarized laser light constant but changing the configuration of the nanoparticles (figure 4(a)). The electron kinetic energy here is chosen based on the synchronicity condition $(\lambda_{ph} v_e)/c = 2r$ [26] to achieve an enhanced coupling constant for each individual nanorod. Therefore, here, we have considered an electron pulse with the kinetic energy of 600 eV interacting with a gold nanorod (15 nm and 10 nm radius, respectively) cavity, illuminated by a laser pulse at the center wavelength of 700 nm and the temporal FWHM broadening of 34 fs.

Therefore, here we have employed three strategies to explore the effect of the plasmonic modes on the electron modulation. In the first arrangement (figure 4, I), where there is an inversion symmetry along the *x*-direction, both nanorods experience the same in-phase strongly-coupled dipolar field (figure 4(a)). In the other configurations, one nanorod is rotated 30°(figure 4, II), and 60°(figure 4, III) around the x-axis, in a way that the projected spacing of all configurations along the y-axis is kept constant. For the first case, only a slight asymmetry is observed in the overall inelastic scattering cross section (Supplementary figure 2 shows more detail) with respect to the origin, while the electron wavepacket receives a total transverse recoil toward the negative y axis.

In the latter two cases, the delay between the wiggling motions induced by each of the dipolar near-field distributions of the individual nanorods (supplementary figure 3) results in an asymmetry in the overall inelastic scattering cross section (figure 4(b)). Particularly, higher-order energy and momentums states are more intensely populated, in contrast with the situation discussed beforehand. For the second configuration, the most intense populated states are centered around $E = 20$ eV and $\varphi = -1°$, as well as $E = -20$ eV and $\varphi = +1°$, whereas for the third case, they are located around $E = -20$ eV and $\varphi = -1°$, as well as $E = +20$ eV and $\varphi = +1°$ ( figure 4(c)).

The distribution of the electron wavepacket along the longitudinal direction spaced by $n$ integers of $\hbar\omega_{ph}$, depicts the critical role of highly confined coupled plasmonic modes for achieving the momentum matching requirement and fostering synchronous motions between near-field oscillations and electron wavepackets. The overall extent of the PINEM spectrum (figure 4(d), are similar for all systems, reaching up to $\pm 40\,\hbar\omega_{ph}$, meaning that quantum interference paths originating from various modes lead to a constructive interference pattern and a strong electron-photon coupling. In addition, since the extent of the gain and loss is comparable with the initial kinetic energy of the electron (13% of the initial kinetic energy), the coupling strength and therefore, the probability amplitude for each loss/gain photon order does not satisfy the nonrecoil approximation[22]. In addition, due to the ultrafast energy modulation in the near-field zone, the accelerated and decelerated parts of the wavepacket experience a different coupling constant with the near-field distribution. The latter phenomenon underpins the observed asymmetry patterns as well.

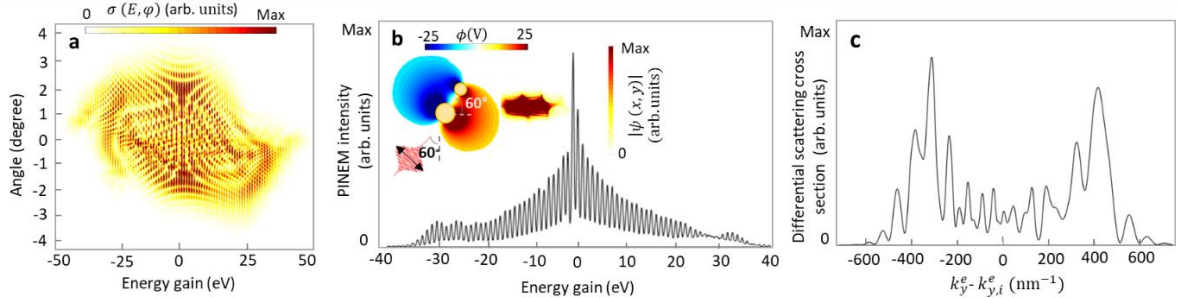

**Figure 5.** Maximizing the transverse recoil with the polarization of the incident light. (a) Inelastic scattering cross section, after the interaction with the localized plasmons in a dimer excited by the laser light with the incident angle of 60°, and (b) the PINEM spectrum, Real-space distribution of the electron wavepacket and the snapshot of the spatial profile of the induced plasmonic near-field at a given time, and (c) transverse recoil of the electron beam along the elastic path $E = 0$. The kinetic energy of the electron beam is 600 eV, and its FWHM longitudinal and transverse broadenings are 45 nm and 3 nm, respectively. The center wavelength of the laser pulse is 700 nm, and its peak field amplitude is 2 GVm$^{-1}$.

When exciting the dimer structure with an inclined laser pulse (figure 5) with its wave vector aligned along the symmetry axis of the dimer, the transverse recoil of the electron beam is also maximized along the elastic path $E = 0$ (figure 5(c)). Interestingly, the overall extent of the wavepacket along the longitudinal direction (figure 5(b)) is still within $-40\,\hbar\omega_{ph}$ to $+40\,\hbar\omega_{ph}$, ascertaining a coupling strength approximately the same as the structures investigated before, regardless of the excitation configuration of the laser beam. Comparing the inelastic scattering cross sections for all the configuration studied here, allows to conclude that while the size of the nanoparticles, the synchronicity condition, and the intensity of the laser field have a large impact on the strength of electron-light interactions, the polarization of the light and the shape of the nanoparticles offer additional degrees of freedoms for tailoring the overall shape of the electron wavepacket along both longitudinal and transverse directions.

## Conclusion

In conclusion, we propose a scheme to shape the electron beam by controlling the spatial distribution of localized plasmonic fields. By controlling the polarization and the geometry of the highly localized optical fields and varying the phase delay between plasmonic fields, the experienced recoil by the electrons is controlled in arbitrary angular deflections. Particularly we investigated the role of electric-field polarization on controlling the wiggling motion of the electron beam interacting with near-field distributions, and thus controlling the overall shape of the electron wavepacket. Flavoring from the merits of the plasmonic cavity for enhanced near-field localization and intensity, we expect our approach may offer a promising platform to study the electron-light interactions on the nanoscale, and it also can be applied to a wide range of applications including ultrafast phase manipulation and free-electron wavepacket shaping for application sin quantum-sensitive measurements.

## Acknowledgements

This project has received funding from the European Research Council (ERC) under the European Union's Horizon 2020 research and innovation program, Grant Agreement No. 802130 (Kiel, NanoBeam) and Grant Agreement No. 101017720 (EBEAM).

## ORCID iDs

Fatemeh Chahshouri https://orcid.org/0000-0001-5920-7805

Nahid Talebi  https://orcid.org/0000-0002-3861-1005

## Supplementary Material:

### S1. Interaction of electron wave packet with single gold nanorod

An electron pulse with the initial center kinetic energy of 200 eV travels in the proximity of a gold nanorod that is excited by a laser pulse and is strongly scattered off the plasmonic near-field. The spatial distribution of the scalar potential depicted in figure S.1 a, shows the direction of the excited dipole. The spatial distribution of the electron wave function after the interaction is demonstrated as well. The angle-resolved scattering cross section of the electron wavefunction in the momentum space (figure S.1(b)) demonstrates that multiple loss and gain peaks occur along the longitudinal axis, whereas several diffraction orders are also observed along the transverse axis. The gradient of the scalar potential along the y-axis exerts a force in the transverse direction on the electron wavepacket, whereas as the time-dependent oscillation of this force causes the electron to be scattered into a wide angular range. In addition though, we observe distinguished Kapitz-Dirac-like diffraction orders, the origins of them could be only described by the quantum-mechanical interferences and be simulated considering the Schrödinger equation. In general, in contrast to the dimer case (figure 3 in the main text), the diffracted peaks here are less pronounced and the transverse diffraction (figure S.1(c)) occurred along the highlighted energy range (demonstrated figure S.1(b)) cannot support sharp resonance peaks. Indeed, the captured near-field in the dimer nanorods adds another degree of freedom to control the transversal recoil and diffraction peaks and is responsible for producing highly separated spectral fringes.

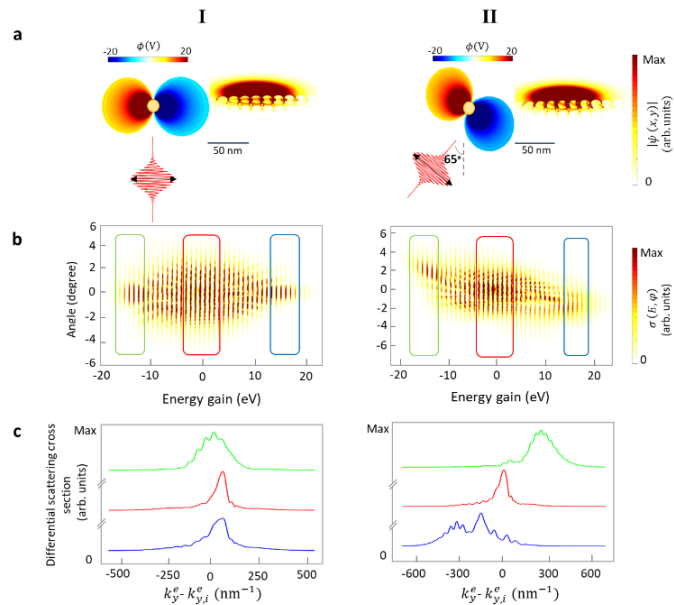

**Figure 1**. Final distribution of the electron wavepacket after the interaction with a gold nanorod excited with a (I) normal, and (II) obliquely polarized laser pulse at the wavelength of 700 nm, and field amplitude of $E_0 = 2 \text{ GVm}^{-1}$. (a) scalar potential, and spatial representation of the electron wavepacket, and (b) Angle resolved energy map of the probability amplitude. (c) Probability distribution of transverse recoil at the highlighted energy range.

## S2. Dynamics of the Interaction and the Electron Recoil

The spatiotemporal behaviour of the electron wavefunction during the interaction with a laser – induced plasmonic dimer is shown in figure S.2 and figure S.3 Electron traveling along the *x*-axis, is keeping its initial energy constant in free space When the electron enters the near-field zone of a larger nanorod at $t = 5.6$ fs, it experiences the longitudinal, as well as transverse Lorentz force. Consequently, it causes a dynamical wiggling motion of the electron wavpacket. An electron moving within the gap experiences more enhanced recoil from the enhanced coupled dipolar mode. Following electron trajectory during the time interval of $t = 5.6$ to 11.3 fs when the electron comes to the middle of the interaction zone, the interference between phase modulation of the first and second dipolar field and the race between them is responsible for shaping the electron recoil and mediating the energy transfer. In this case, the distance between these two nanorods and the delay between the optical excitations in them affects the modulation of the electron wavepacket. These two structures show how the relative phase delay can enhance or hamper the already occupied longitudinal and transverse momentum orders. For example, in the case when second nanorod is rotated $30°$ (figure S.3(b)), electron experiences an overlapped plasmonic fields of both nanorods in a longer nearfield zone. Thus, it has longer interaction time with the fields. Whereas, for the case where the second nanorod is placed on top of the first one (figure S.2(b)), the electron feel both near-field zone at the same time. Hence, the phase delay in the excitation of the nanorods can control subsequent constructive or destructive electron-photon interactions and the final scattering process.

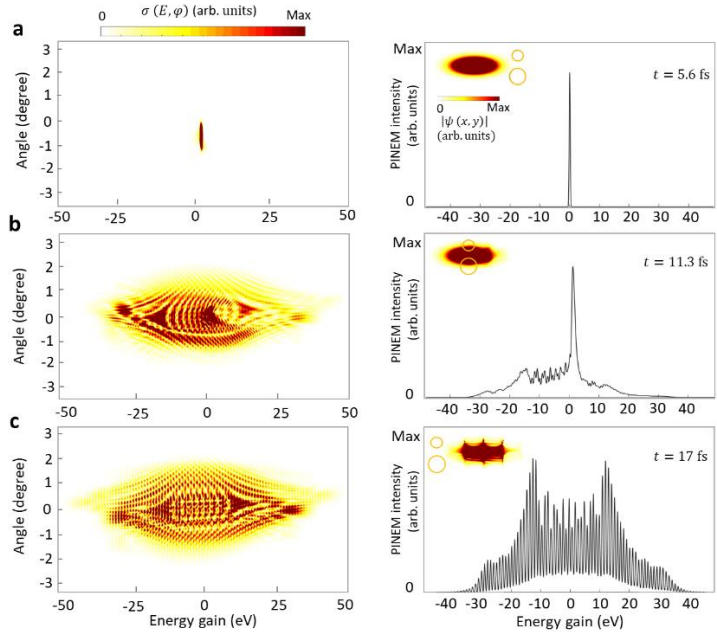

**Figure 2.** Temporal evolution of the electron wavepacket during its interaction with the enhanced near-field excited with an *x*-polarized laser pulse at the wavelength of 700 nm, and the peak field amplitude of $E_0 = 2$ GVm$^{-1}$. (a) Evolution of the electron wavepacket in the reciprocal space and (b) the energy- gain spectra by integrating over the entire angular distribution at a given time during the interaction. The gold nanorods radius is 15 nm and 10 nm respectively with 20 nm gap spacing.

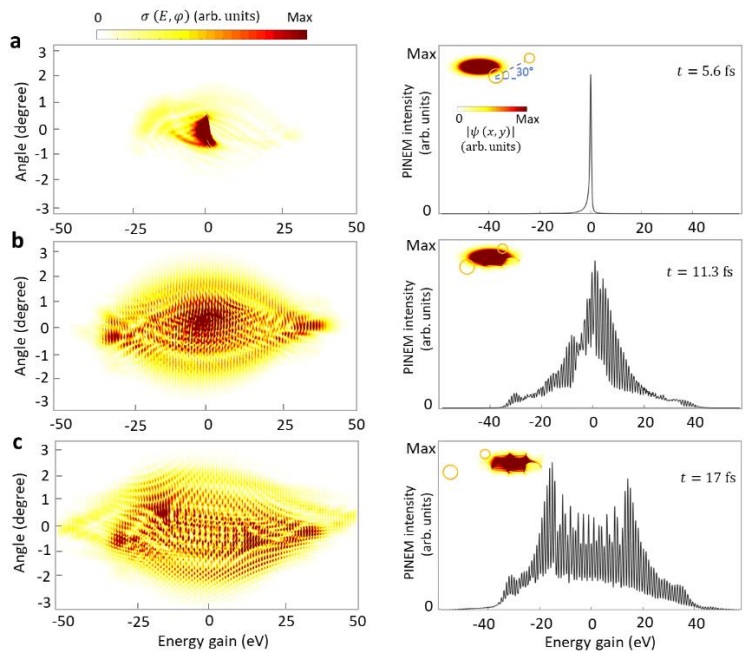

**Figure 3.** Dynamics of the experienced recoil of the electron wavepacket during its interaction with localized plasmonic field excited with an *x*-polarized laser pulse at the wavelength of 700 nm, field amplitude of $E_0 = 2$ GVm$^{-1}$. (a) The electron wavepacket in the momentum space, and (b) the energy gain spectra calculated during the interaction at a given time. Inset depicts the evolution of the electron wavepacket in real space at given times.